\documentstyle{mn}

\newif\ifAMStwofonts
\ifoldfss
  \ifCUPmtlplainloaded \else
    \NewTextAlphabet{textbfit} {cmbxti10} {}
    \NewTextAlphabet{textbfss} {cmssbx10} {}
    \NewMathAlphabet{mathbfit} {cmbxti10} {} % for math mode
    \NewMathAlphabet{mathbfss} {cmssbx10} {} %  "   "    "
  \fi
  \ifAMStwofonts
    \ifCUPmtlplainloaded \else
      \NewSymbolFont{upmath} {eurm10}
      \NewSymbolFont{AMSa} {msam10}
      \NewMathSymbol{\upi}     {0}{upmath}{19}
      \NewMathSymbol{\umu}     {0}{upmath}{16}
      \NewMathSymbol{\upartial}{0}{upmath}{40}
      \NewMathSymbol{\leqslant}{3}{AMSa}{36}
      \NewMathSymbol{\geqslant}{3}{AMSa}{3E}

    \fi
  \fi
\fi % End of OFSS

\ifnfssone
  \newmathalphabet{\mathit}
  \addtoversion{normal}{\mathit}{cmr}{m}{it}
  \addtoversion{bold}{\mathit}{cmr}{bx}{it}
  \newmathalphabet{\mathbfit} % math mode version of \textbfit{..}
  \addtoversion{normal}{\mathbfit}{cmr}{bx}{it}
  \addtoversion{bold}{\mathbfit}{cmr}{bx}{it}
  \newmathalphabet{\mathbfss} % math mode version of \textbfss{..}
  \addtoversion{normal}{\mathbfss}{cmss}{bx}{n}
  \addtoversion{bold}{\mathbfss}{cmss}{bx}{n}
  \ifAMStwofonts
    \ifCUPmtlplainloaded \else
      \UseAMStwoboldmath
      \makeatletter
      \new@mathgroup\upmath@group
      \define@mathgroup\mv@normal\upmath@group{eur}{m}{n}
      \define@mathgroup\mv@bold\upmath@group{eur}{b}{n}
      \edef\UPM{\hexnumber\upmath@group}
      \new@mathgroup\amsa@group
      \define@mathgroup\mv@normal\amsa@group{msa}{m}{n}
      \define@mathgroup\mv@bold\amsa@group{msa}{m}{n}
      \edef\AMSa{\hexnumber\amsa@group}
      \makeatother
      \mathchardef\upi="0\UPM19
      \mathchardef\umu="0\UPM16
      \mathchardef\upartial="0\UPM40
      \mathchardef\leqslant="3\AMSa36
      \mathchardef\geqslant="3\AMSa3E
    \fi
  \fi
\fi % End of NFSS release 1

\ifnfsstwo
  \DeclareMathAlphabet{\mathbfit}{OT1}{cmr}{bx}{it}
  \SetMathAlphabet\mathbfit{bold}{OT1}{cmr}{bx}{it}
  \DeclareMathAlphabet{\mathbfss}{OT1}{cmss}{bx}{n}
  \SetMathAlphabet\mathbfss{bold}{OT1}{cmss}{bx}{n}
  \ifAMStwofonts
    \ifCUPmtlplainloaded \else
      \DeclareSymbolFont{UPM}{U}{eur}{m}{n}
      \SetSymbolFont{UPM}{bold}{U}{eur}{b}{n}
      \DeclareSymbolFont{AMSa}{U}{msa}{m}{n}
      \DeclareMathSymbol{\upi}{0}{UPM}{"19}
      \DeclareMathSymbol{\umu}{0}{UPM}{"16}
      \DeclareMathSymbol{\upartial}{0}{UPM}{"40}
      \DeclareMathSymbol{\leqslant}{3}{AMSa}{"36}
      \DeclareMathSymbol{\geqslant}{3}{AMSa}{"3E}
    \fi
  \fi
\fi % End of NFSS release 2

\ifCUPmtlplainloaded \else
  \ifAMStwofonts \else % If no AMS fonts
    \def\upi{\pi}
    \def\umu{\mu}
    \def\upartial{\partial}
  \fi
\fi

\title{RXTE/OSSE Fits to the Hard State Spectrum of Cygnus X-1}
\author[T. J. Maccarone \& P. S. Coppi]
       {Thomas J. Maccarone \\
        Scuola Internationale Superiore di Studi Avanzati, via Beirut, n. 2-4, Trieste, Italy, 34014 
	\newauthor
	Paolo S. Coppi\\
	Department of Astronomy, Yale University, P.O. Box 208101, New Haven CT USA 06520-8101}

\date{}

\pagerange{\pageref{firstpage}--\pageref{lastpage}}
\pubyear{}

\begin{document}

\maketitle

\label{firstpage}
\input{epsf}
\input{epsfig.sty}

%\label{firstpage}

\begin{abstract}

We present three spectra of Cygnus X-1 in the low/hard state from
3-1000 keV produced with RXTE and OSSE.  We demonstrate that a pure
thermal Comptonization model, with Compton reflection fits the data
well, with coronal temperatures of about 90 keV, optical depths of
about 1.3, and reflection fractions of about 0.25 $\times$ 2$\pi$.  We
find that no additional components are necessary to fit the data,
although the presence of an additional non-thermal component does
result in a marginal improvement to the goodness of the fit.  Two of
the observations are fit better by an almost purely non-thermal
Comptonization model with $\Gamma_{inj}$ of $\sim$ 3.2 for the
electrons and the other parameters essentially the same as for the
thermal model, while the third observation is also consistent with
this model.  Observations that could break this degeneracy are
discussed.  We also demonstrate that the low reflection parameter
(i.e. $R/2\pi<1$) is due to a lack of strong curvature in the 30-50
keV range of the spectrum and not due to ionization effects.  The
spectrum changes very little over the three weeks of the observations,
in accordance with a three week correlation timescale found in the All
Sky Monitor data.  We show that for a purely thermal corona, the
radius of the corona must be between $3\times 10^6$ and $1.5\times
10^{10}$ cm, while the radius of a hybrid corona or a purely
non-thermal corona is constrained to be between $10^8$ and $10^9$ cm.

\end{abstract} 


\section{Introduction} 

Cygnus X-1 is a binary star system consisting of an O9.7 Iab
supergiant (Gies \& Bolton 1986) and a compact object of at least 6
solar masses (Dolan 1992).  The compact object accretes matter from a
focused wind off the supergiant star (Gies \& Bolton 1986) and is one
of the brightest persistent sources of X-rays in the sky.  Cyg X-1 is
the first system to show strong evidence of containing a black hole
and remains one of the strongest candidates for being a black hole.

Phenomenologically, the spectrum of Cygnus X-1 above $\sim$ 3 keV is
well described as an exponentially cutoff power law with a photon
spectral index of $\Gamma \sim 0.6-0.7$ and a cutoff energy of around
150 keV.  Additionally the spectrum shows a broad excess around 30-50
keV due to Compton reflection, an iron K edge above 7 keV and an iron
K$\alpha$ flourescence emission line (see e.g., Ebisawa 1996).  The
hard power law has generally been described as arising from thermal
Comptonization of soft photons in a hot, photon starved corona
(Sunyaev \& Tr\"umper 1979; Gierlinski et al. 1997; Poutanen, Krolik,
\& Ryde 1997; Dove et al. 1998).

A previous attempt to model the X/$\gamma$ spectrum of Cygnus X-1
relied on data from Ginga and OSSE (Gierlinski et al., 1997).  They
showed that a thermal Comptonization model with $kT \sim 100$ keV,
$\tau\sim1$ plasma could fit the data, but argued that patterns found
in the residual plot suggest a two component model with a hot moderate
optical depth ($\tau \sim 1-2, kT \sim 100$ keV) plasma and an
additional Wien-like component with $kT \sim 50$ keV and $\tau \sim
6$.  An attempt to model the broadband X-ray spectrum from 3 to 200
keV showed that the data could be fit about equally well by a variety
of models - thermal Comptonization with a coronal temperature of 40
keV and an optical depth of 3.6, a different Comptonization model with
a temperature of 87 keV and an optical depth of 1.6, or a cutoff power
law with an index of 1.45 and a cutoff energy of 160 keV, plus an
additional blackbody component, but several of the relatively good
(i.e. $\chi^2/\nu < 2$) fits suggested a covering fraction for Compton
reflection consistent with zero (Dove et al., 1998).  What the past
analyses lack and these analyses provide is full broadband coverage
from 3 to 1000 keV.  The combination of Ginga and OSSE leaves a gap
from roughly 30 to 50 keV where the reflection component most affects
the broadband spectrum while the use of RXTE alone fails to reach the
energies where the spectrum cuts off most steeply at $\sim$ 400 keV.
Here we present the results of joint RXTE/OSSE fits to the spectrum of
Cygnus X-1 taken in December of 1998.  These data present the first
gap-free energy spectra of Cygnus X-1 from 3 to 1000 keV.

\section{OBSERVATIONS} 

We examine data from three OSSE viewing periods where there was
simultaneous RXTE data.  The color changes measured in the PCA and in
the ASM over the one week periods are very small (less than a few
percent; see Figure 1a for the PCA colors), suggesting that the
spectrum remains stable over these time periods.  Also, the ASM summed
band intensities (1.3-12.0 keV) show a correlation timescale of about
20 days (see Figure 1b; the 1996 data have been excluded so the
measure will not be biased by the state transition that occured then).
The OSSE fluxes in the highest energy bands are weak and only
detectors 3 and 4 from OSSE were turned on.  Hence the full OSSE
viewing periods, rather than just the strictly simultaneous data, are
used as the average spectra for the analysis.  The data are extracted
using the standard reduction procedure from the CGRO science center.

\begin{figure*}
\centerline{\psfig{figure=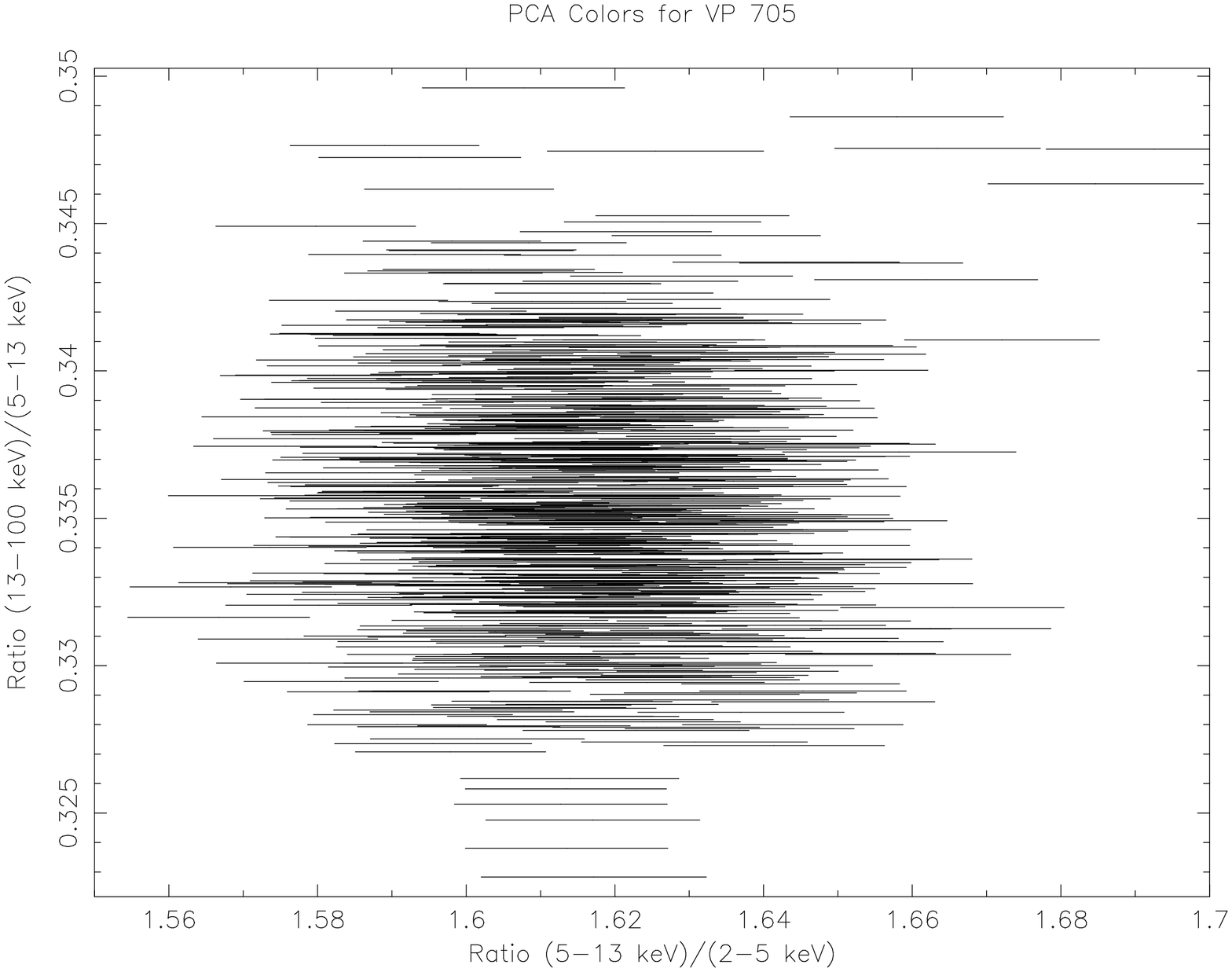,width=9cm,angle=-90}\psfig{figure=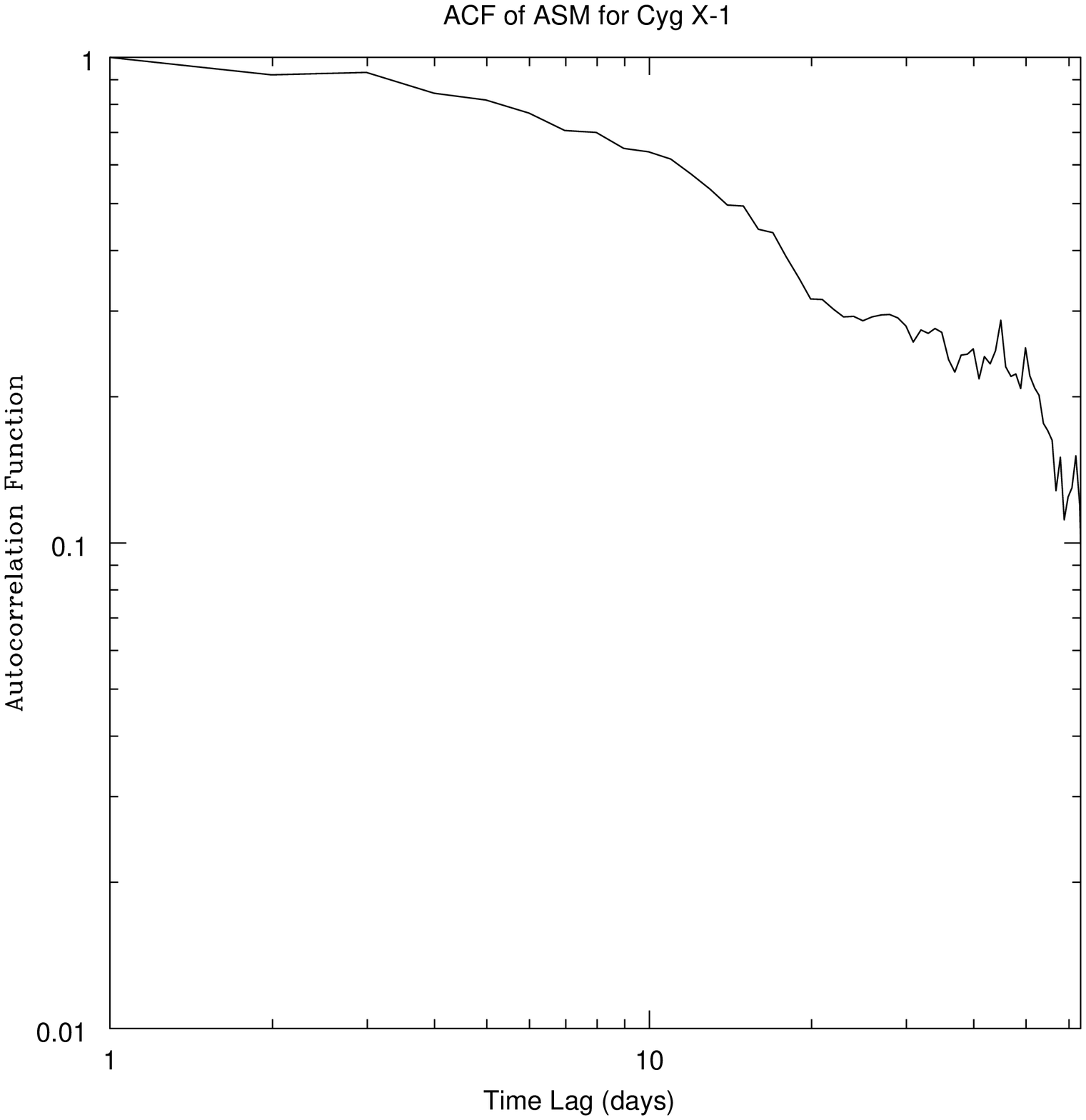,width=9cm}}
\caption{(a)The PCA colors over the first week of the observation.
The colors change very little within the observations and by only a
few percent from observation to observation. A few points are omitted
from the plot near zero orbital phase where nH variations make a large
difference in the colors and count rates (b) The autocorrelation
function of the summed band (1.3-12.0 keV) ASM lightcurve.}
\end{figure*}   

The RXTE data are analyzed using the standard RXTE GOF criteria for
spectral extraction.  Data from only the top layers are used, and all
five PCUs are added together.  The HEXTE data are also extracted using
the standard reduction criteria.  The data used from each instrument
is presented in Table 1.

\section{ANALYSIS}

The data are then fit with the EQPAIR model using XSPEC 11.0 (Arnaud
1996).  The model allows for thermal and non-thermal comptonization of
soft photons, pair production, Compton reflection, and bremsstrahlung
emission.  The physics of the model are described in detail in Coppi
(1998).  It has previously been used to fit spectra from Cyg X-1
(Gierlinski et al. 1997; Frontera et al. 2001), GX 339-4 (Nowak, Wilms
\& Dove 2002) and GRS 1915+105 (Maccarone, Coppi \& Taam 1999;
Maccarone 2001; Zdziarski et al. 2001).  One percent systematic errors
are added to all channels.  Many of the parameters are frozen.  The
inclination angle is assumed to be 55 degrees, in agreement with the
most recent measurements (Sowers et al, 1998).  The compactness,
$\ell_{bb}$ ($\ell_{bb} = L_{bb}\sigma_T/m_ec^3R_{cor}$, where
$L_{bb}$ is the luminosity of the seed blackbody photon distribution,
$\sigma_T$ is the Thomson cross-section, $m_e$ is the electron mass,
and $R_{cor}$ is the radius of the emission region) of the seed photon
distribution is assumed to be unity.  The elemental abundances are
assumed to be solar.  The reflector's temperature is assumed to be 100
eV.  The overall normalizations of the HEXTE and OSSE components are
allowed to float, but are extremely well constrained since even small
deviations (i.e. $\sim$ 1\%) from the best fit normalizations will
produce extremely statisically significant discontinuities in the
spectra.  The PCA data are fit from 3 to 20 keV, the HEXTE from 17 to
190 keV and the OSSE from 100 to 1000 keV.  Additional model
components of neutral hydrogen column absorption and a gaussian
component to fit the iron emission line are also included.  Since RXTE
is not sensitive to the energies where absorption by the interstellar
medium is most important, the neutral hydrogen column is fixed to $5
\times 10^{21}$ in accordance with ASCA and BeppoSax results
(Gierlinski et al. 1999; Frontera et al. 2001 ).  The iron line is
restricted to fall between 6.1 and 7.0 keV in energy and to be less
than 1.2 keV in width.  The parameters that are allowed to vary freely
are the temperature of the seed blackbody photon distribution
($kT_{bb}$), the thermal electron compactness ($\ell_{th}$), the
reflection fraction, $R$ and the ionization parameter $\xi$, as well
as an overall normalization.  Additionally, the HEXTE and OSSE data
sets are allowed to have their normalizations float freely since the
dead time correction for the HEXTE is uncertain (especially with the
standard RXTE GOF distribution response matrices which we have used)
as are the cross-calibrations between RXTE and OSSE.  Additionally,
since the data from the two satellites are not strictly simultaneous
there may be some physical difference in the luminosities apart from
calibration issues.  Typically, the HEXTE and the OSSE are both
normalized to about 70\% of the PCA.  We also note that these
observations are typical for Cygnus X-1 as the power spectrum shows
breaks at $\sim$ 0.1 Hz and $\sim$ 5 Hz, which are approximately the
median values for the break frequencies (Pottschmidt et al., 2001).
Since the spectral indices tend to change much less than the
variability timescales (see e.g. Belloni \& Hasinger 1990), and the
two are well correlated, this represents strong evidence that the
observations are typical.

\subsection{Pure thermal Comptonization model fits}

The results of the fitting are presented in Table 2.  A typical
spectral fit is plotted in Figure 2.  The residuals to this fit are
shown in Figure 3.  Most of the parameters vary little from
observation to observation, in accord with the result that the colors
in the PCA bands vary little over this time span.  The parameters that
change the most are the iron line equivalent width, the ionization
parameter for the reflection, and the and the iron line's physical
width.  Since these parameters are all most sensitive to the region
between 6 and 8 keV (the ionization level parameter depending
primarily on the energy of the iron absorption edge), where there are
not sufficient independent energy bins to fit the required number of
parameters, one cannot be confident that the the differences are real.
Furthermore, the best fit equivalent widths of the iron lines are of
order 100 eV for these observations.  Previous measurements of the
iron line for Cygnus X-1 in the hard state from ASCA show $\sim$ 20 eV
equivalent width (Ebisawa et al. 1996).  RXTE fits to the Crab
spectrum can suggest the presence of an iron line with an equivalent
width of about 80 eV, so features this weak should not be trusted
regardless of the formal statistical significance of the result.
Errors in the iron line measurements can lead to significant errors in
the edge position.  That the line energies are found to be below 6.4
keV at a strong signficance level (in disagreement with results from
ASCA) also indicates that the narrow feature is more likely related to
errors in the response matrix than to a real physical feature.

\begin{figure*}
\centerline{\psfig{figure=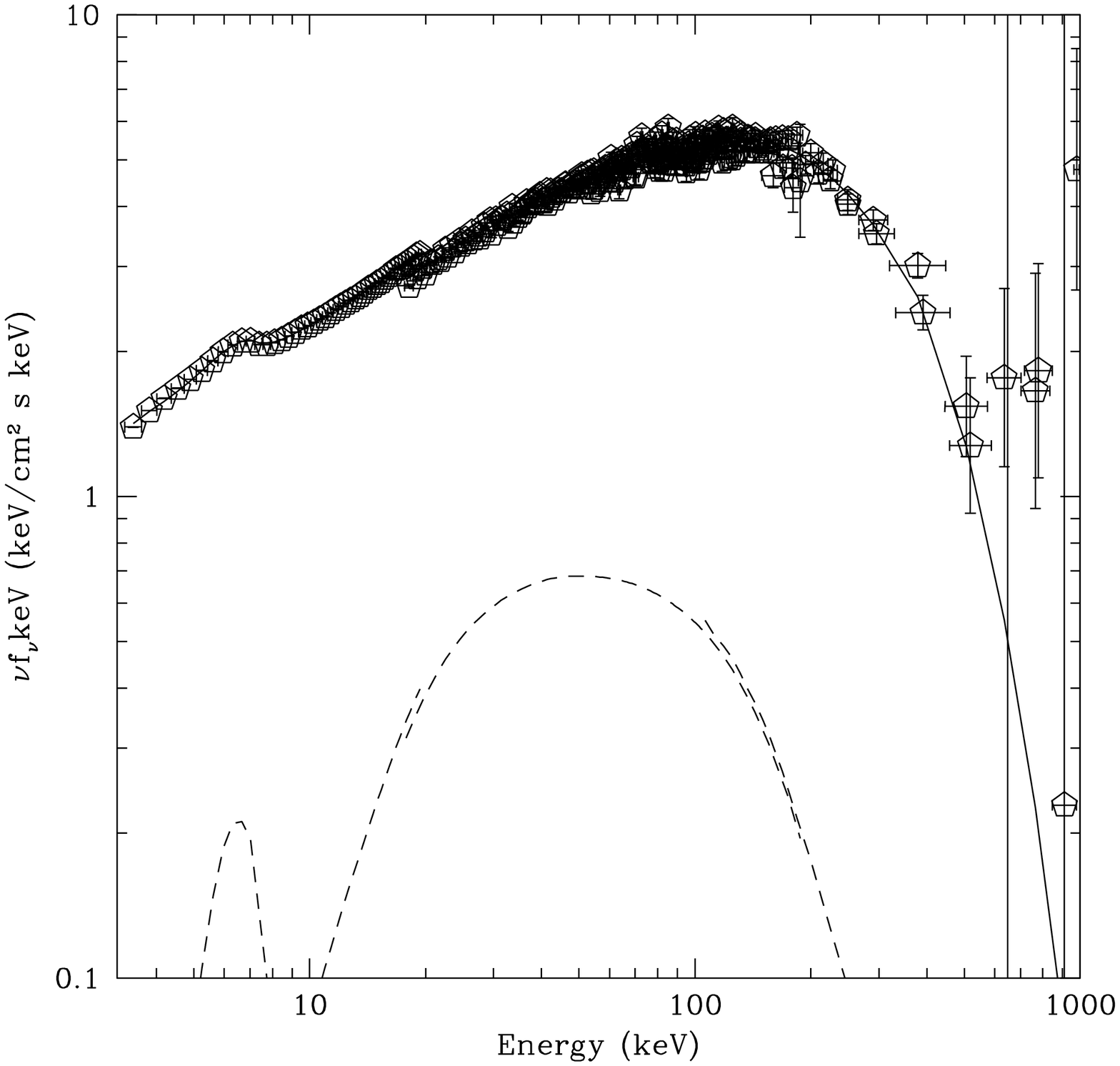,width=18cm}}
\caption{The unfolded thermal Comptonization fit to the data for OSSE
viewing period 707, plotted in $E^2\frac{dN}{dE}$ units.  The data are
in open pentagons, the model fits in the solid line and the reflection
component in the dashed line.  The PCA component has been multiplied
by 0.7 to make its relative normalization approximately the same as
those for HEXTE and OSSE.}
\end{figure*}   

\begin{figure*}
\centerline{\psfig{figure=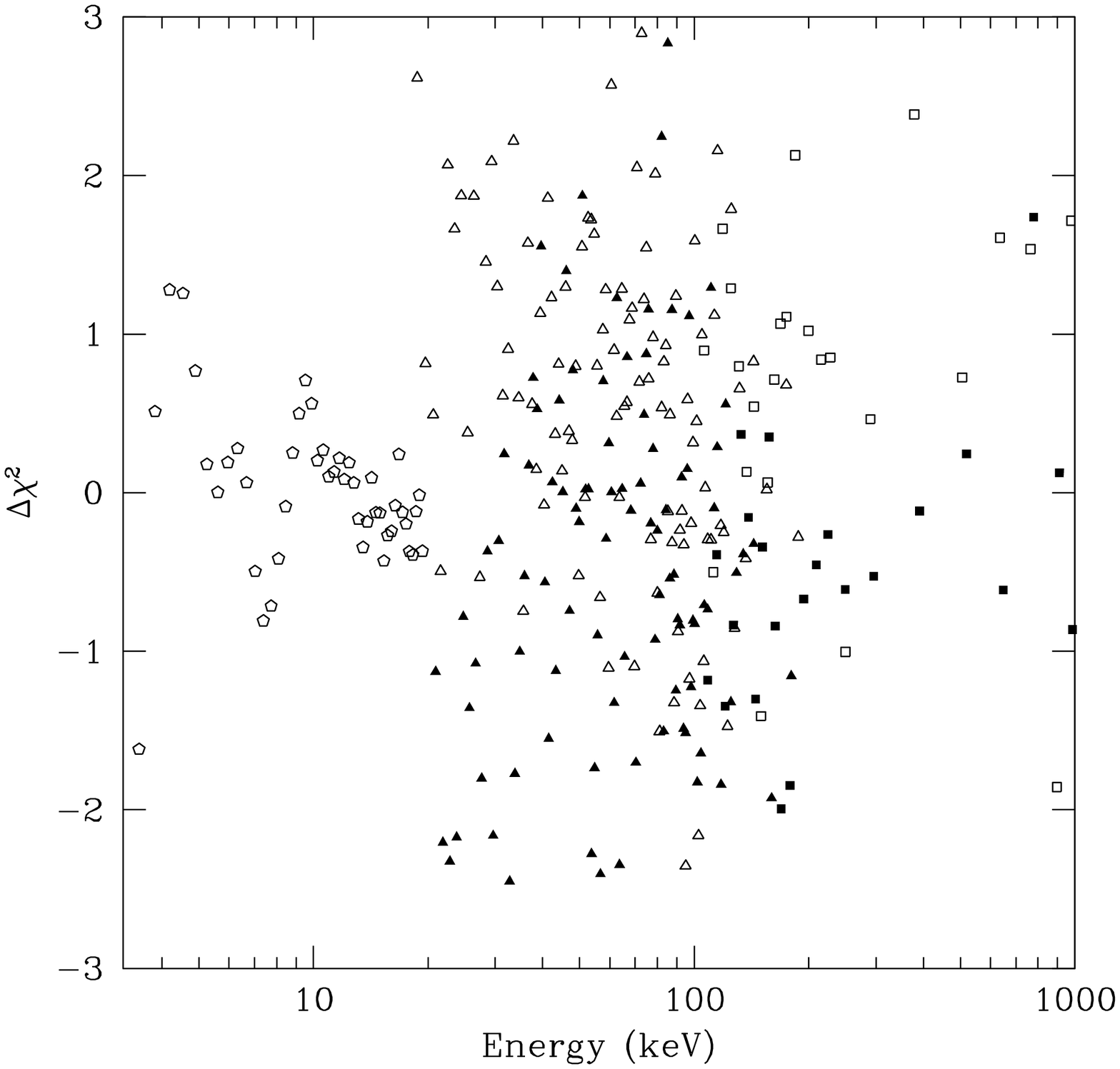,width=18cm}}
\caption{The residuals to the fit for OSSE viewing period 707, plotted
as $\Delta\chi^2$ versus energy.  The PCA points are open pentagons,
the HEXTE cluster 0 points are open triangles, the HEXTE cluster 1
points are filled triangles, the OSSE cluster 3 points are open
squares and the OSSE cluster 4 points are filled squares.}
\end{figure*}

Other than the line parameters, a clear picture for the spectrum
emerges where the disk temperature is fit to be between 150 and 250
eV, the ratio of thermal compactness to seed photon compactness is
about 9, and the optical depth is about 1.3.  Thus previous results
that the corona of Cygnus X-1 in the hard state can be well fit as a
photon-starved thermal plasma of moderate optical depth are verified.
We have done several simulations and found that changes in the
blackbody temperature can be measured due to changes in the curvature
of the ``power law'' component of the spectrum at energies well in
excess of the blackbody peak.  For these observations, the blackbody
temperatures vary very little; this model would be sensitive to
variations about twice as large.  Since slightly different models for
the blackbody seed photon distribution can change the observed
temperature and because of uncertainties in the RXTE response matrix
and the lack of an RXTE response below 2 keV where the blackbody
peaks, {\it absolute} measurements of the seed photon temperature
should be treated with skepticism.  Relative changes, however, can be
reliably measured.

\subsection{Hybrid Thermal/Non-thermal Comptonization model fits}
Additional fits are made with the inclusion of a non-thermal component
to the spectrum.  That is to say, the electron distribution then
consists of a Maxwellian distribution (the ``thermal'' Comptonization
component) and a power law component (the ``non-thermal''
Comptonization component, which is restricted to have electron Lorentz
factors from 1.3 to 1000 and has a compactness $\ell_{nth}$ and an
electron spectral index of $\Gamma$ such that $\frac{dN}{dE} \sim
E^{-\Gamma}$). The value of $\Gamma$ is required to be greater than 1,
the hardest electron spectrum that can be produced by standard
acceleration processes.  The best fit parameters are shown in Table 3.
These fits are marginally better than those without the non-thermal
component, with improvements in likelihood of factors of about 5 for
VP 705 and VP 706 (which are best fit by a nearly purely {\it
non-thermal} Comptonization model), and no improvement in VP 707.
Since it is unlikely that a major qualitative change (i.e. whether the
corona is dominated by thermal or non-thermal electrons) has occurred
while the spectra look almost identical, we also attempt to fit a
model to the data with a purely non-thermal electron for VP 707.  We
find that the data are consistent with this model as well
($\chi^2/\nu=1.04$), with $\Gamma_{inj}$=3.3, $\ell_{nth}$=8.5, and the
other paraemters consistent with their values for both the pure
thermal and the hybrid model cases.

The difficulty of distinguishing a thermal plasma from a hybrid one
has been discussed previously.  It has been noted that a hybrid plasma
and a purely thermal plasma will produce nearly identical spectra
given the same optical depth and the same mean energy shift per
scattering (Zdziarski, Coppi, \& Lamb 1990; Ghisellini, Haardt, \&
Fabian 1993).  Still theoretical models do predict differences between
the two plasmas above $\sim$ 400 keV (see Figure 4), since the purely
thermal Comptonization models will show a nearly exponential decay,
while the hybrid models will show a power law decay.  Distinguishing
between these two best fits would require sensitivity at the
$4\times10^{-6}$ photons/cm$^2$/sec level, which is beyond the
capability of the existing OSSE observations.  A 10-$\sigma$ detection
of this feature should be feasible with INTEGRAL in $\sim 10^6$
seconds, and with ASTRO-E2 in about $10^5$ seconds.  Measurements from
COMPTEL also seem to suggest the presence of an extended tail to the
spectrum (McConnell et al., 2000), but these observations are not
taken simultaneously with X-ray detections and also require long
integrations over which variability of the coronal temperature could
be important.

\begin{figure*}
\centerline{\psfig{figure=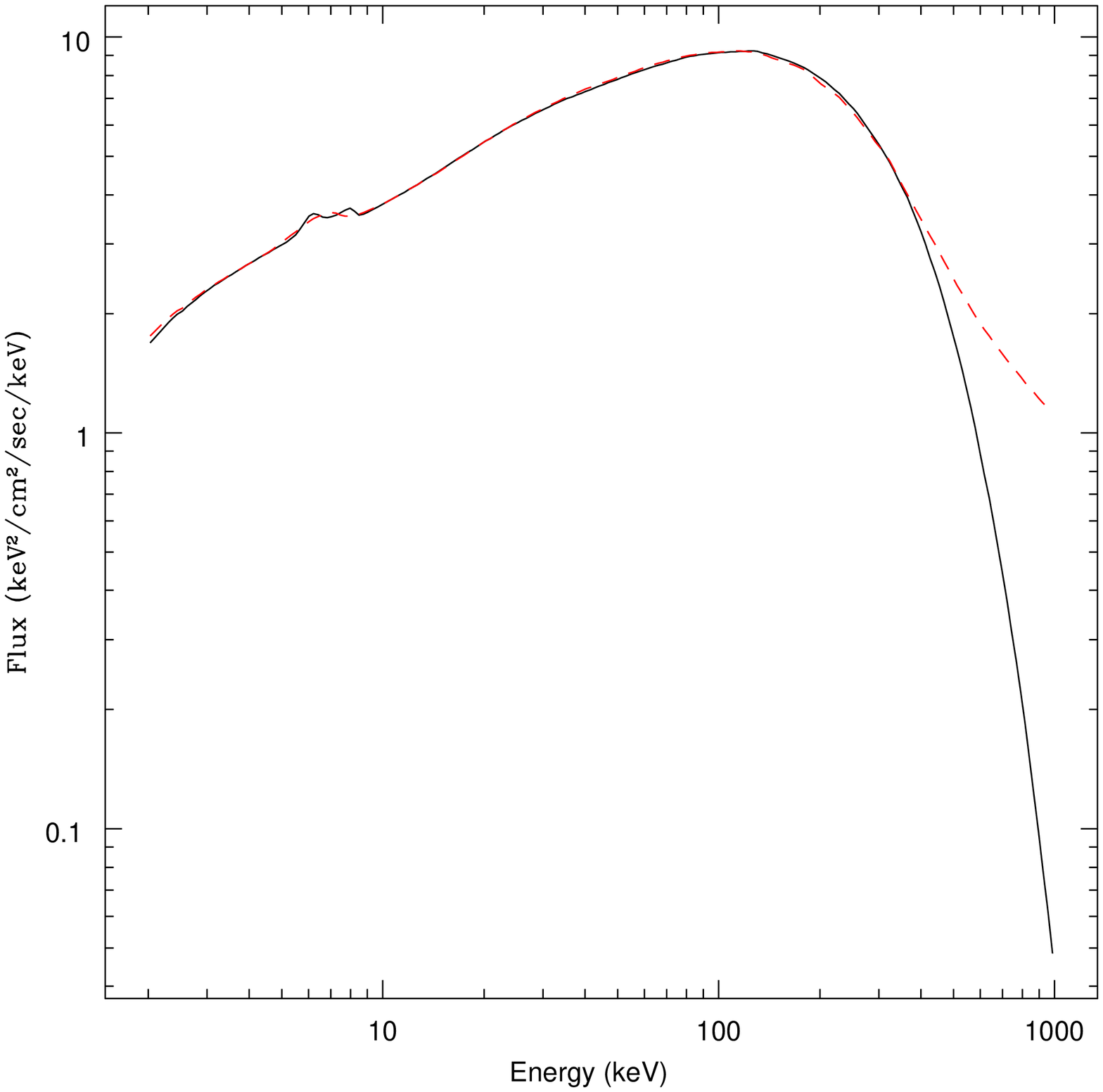,width=18cm}}
\caption{The plot of the thermal and non-thermal fits to the spectrum
from VP 706, folded through an idealized response matrix.  Note that the differences are only apparent above $\sim$ 400 keV.}
\end{figure*}   

\subsection{The Overall Compactness and the Size of the Emission Region}
We also investigate the range of overall compactnesses allowed by the
data to determine whether constraints can be placed on the radius of
the emission region and on the fraction of the coronal optical depth
that is $e^-/e^+$ pairs.  We re-fit the model for VP 707 with a purely
thermal electron distribution, while varying $\ell_{bb}$.  We find at
the 90\% confidence level that the total compactness of the system
cannot exceed $\sim$ 300.  This constraint requires that $R_{cor} > 3
\times 10^6$ cm, which is not a strong constraint since this value is
$\sim 1 R_{SCH}$ for a 10 $M_{\odot}$ black hole.  We also find that
the total compactness cannot drop below $\sim$ .06 or the
bremsstrahlung cooling rate will dominate the Compton cooling rate,
yielding a bremsstrahlung emission spectrum which has too sharp a
cutoff at high energies.  This constraint requires that $R_{cor}< 1.5
\times 10^{10}$ cm.  This provides a spectral confirmation that the
$\sim$ few second hard time lags seen in variability analyses cannot
be produced by light travel times in a large corona as had been
suggested previously (see e.g. Kazanas, Hua, \& Titarchuk 1997) and
refuted on the basis of variability measurements (Maccarone, Coppi, \&
Poutanen 2000).

We conduct the same analysis to find the minimum and maximum
compactness for the purely non-thermal spectral fit to VP 707.  The
best fit is for $\ell_{bb} = 1.5\pm^{2.5}_{0.3}$.  The tighter bound
in the lower direction is because there is some intrinsic curvature to
the spectrum and some $\gamma-\gamma$ pair production is needed to
reproduce this.  The tighter bound in the upward direction is because
the annihilation peak at 511 keV is much sharper for non-thermal
Comptonization than for thermal Comptonization (i.e. in the thermal
Comptonization case, the annihilation line is reddened and broadened
significantly by Compton scattering off the thermal electrons - see
Coppi 1998).  These constraints require a corona with a size scale
between $2\times10^8$ and $7\times10^8$ cm.  This result is not very
sensitive to letting $\gamma_{min}$ float freely, so if a purely
non-thermal electron distribution can be proven to exist, fairly
strong constrints can be placed on the size of the emission region.  A
hybrid model (rather than purely non-thermal) produces similarly tight
constraints as the purely non-thermal model ($\ell_{bb} =
2.2\pm_{.5}^{1.5}$).

\subsection{The reflection fraction}
The relatively low reflection fractions in Cyg X-1 and other X-ray
binaries have been explained as coming from a geometric effect where
there is a ``hole'' inside the inner edge of the accretion disk from
which no reflection comes (Gilfanov et al., 1999), a beaming effect
due to a mildly relativistic outflow from the disk (Beloborodov 1999),
smearing of the reflected component as it passes back through the
corona (Petrucci et al., 2001) or an underestimate of the reflection
fraction due to a highly ionized disk (Young et al. 2001).  We test
the last of these mechanisms on VP 707 by allowing freezing the
ionization parameter at a value of $10^5$.  We find that even for this
large ionization parameter, the reflection fraction remains $\sim$
0.35.  Furthermore, the spectral fits are substantially poorer
($\Delta\chi^2 \sim 150$) than the best fits, which are obtained with
low ionization parameters.  We also ignore the PCA data from 5.0 to
9.5 keV and re-fit the model to ensure that the iron line and edge
have no impact on the best fit parameters and find that none of the
parameters change.  The reason for the discrepancy between this work
and that of Young et al. (2000) is likely that the reflection fraction
in the ASCA/Ginga data sets they fit depends largely on the location
and strengths of the iron edges since ASCA has no sensitivity above 10
keV and Ginga has none above 30 keV, while in our fits, the reflection
fraction depends largely on the curvature of the spectrum in the 30-50
keV range, since there are many more bins in this region than the 6-9
keV region where the iron line and edge are found.  We have also fit
the data after removing the OSSE data and have found essentially the
same fit as for the PCA and HEXTE data alone.

\subsection{Robustness of the fits to changes in the frozen parameters}
We investigate the robustness of the spectral fits to changes of
several parameters that have been left frozen.  We find that changing
the seed photon distribution to a diskPN model, which gives the sum of
several blackbodies as from a pseudo-Newtonian disk (Gierlinski et
al., 1999), rather than a pure blackbody yields a slightly higher
optical depth ($\tau\sim1.4$) and a higher thermal compactness
($\ell_{th}\sim12$) without making any qualitative changes to the
spectrum or the overall goodness of fit.  We find that for the
nonthermal model, $\gamma_{min}$ can be increased to $\sim3$ without
affecting the quality of the fit.  For such high values of
$\gamma_{min}$, the electron injection distribution approaches a
$\delta$-function, so these fits are probably unphysical, and it is
unlikely that the non-thermal electron distribution is cut off much
below 200 keV.

\section{Conclusions}

We find that the 3-1000 keV spectrum of Cygnus X-1 in the hard state
can be well fit by a purely thermal Comptonization model, but that a
substantial (and perhaps dominant) non-thermal component cannot be
ruled out.  For a thermal model, we find typical temperatures of the
corona to be $\sim$ 90 keV, with optical depths of $\sim$ 1.3, and
that for two of the observations, the data can be modelled slightly
better with a purely nonthermal electron distribution.  We find no
evidence for the existence of an additional optically thick component
in the spectrum.  We find that the data are insufficient to
differentiate between a pair-domianted corona and an electron/ion
dominated corona.  We find that the reflection fractions are low
(i.e. $R/2\pi<1$) because there is relatively little spectral
curvature in the 30-50 keV range and not due to a ionization effects
in the disk.  Future observations with instruments such as INTEGRAL
and ASTRO-E2 should prove capable of breaking the degeneracy between
the thermal and non-thermal electron distributions by measuring the
emission at several MeV more accurately than current data sets which
have provided simultaneous X-ray measurements.

\section{Acknowledgments}

We wish to thank J\"orn Wilms and Mike Nowak for scripts that
facilitated finding the errors in the spectral fits.

\begin{table*}
\begin{center}
{\bf Observations Analyzed for This Work}

\begin{tabular}{ccc}
{\bf ObsID} & {\bf Start Time} & {\bf Stop Time} \\

30158-01-01-00 & 10/12/97 07:07:53 & 10/12/97 08:30:14\\
30158-01-02-00 & 11/12/97 07:06:14 & 11/12/97 08:45:14\\
30158-01-03-00 & 14/12/97 08:48:14 & 14/12/97 10:20:14\\
30158-01-04-00 & 15/12/97 03:49:42 & 15/12/97 05:26:14\\
30158-01-05-00 & 15/12/97 05:26:14 & 15/12/97 07:09:14\\
30158-01-06-00 & 17/12/97 00:39:55 & 17/12/97 02:05:14\\
30158-01-07-00 & 20/12/97 07:11:18 & 20/12/97 08:29:14\\
30158-01-08-00 & 21/12/97 05:28:14 & 21/12/97 07:05:14\\
30158-01-09-00 & 24/12/97 23:03:14 & 24/12/97 00:39:14\\
30158-01-10-00 & 25/12/97 00:39:14 & 25/12/97 01:45:14\\
30158-01-11-00 & 30/12/97 02:18:03 & 30/12/97 03:52:14\\
30158-01-12-00 & 30/12/97 03:52:14 & 30/12/97 05:30:14\\

{\bf Viewing Period} & {\bf Start Time} & {\bf Stop Time}\\

705 & 1997-12-09 17:44:24 & 1997-12-16 14:08:23 \\
706 & 1997-12-16 16:54:51 & 1997-12-23 13:27:23 \\
707 & 1997-12-23 16:18:54 & 1997-12-30 14:15:46 \\

\end{tabular}
\end{center}
\caption{The RXTE and OSSE observation times.  VP 705 in OSSE
coincides with RXTE observations 30158-01-01-00 through
30158-01-05-00, VP 706 coincides with 30158-01-06-00 through
30158-01-08-00 and VP 707 coincides with 30158-01-09-00 through
30158-01-12-00.}

\end{table*}

\begin{table*}
\begin{center}
{\bf Best Fit Parameters for the Thermal Model}

\begin{tabular}{cccccccccccc}
VP & $kT_{bb}$ & $\ell_{th}$ & $\tau$ & $R$ & $\xi$ & $F$ & LE & LW & LN & EqW & $\chi^2/\nu$ \\

705 & $151\pm^{29}_{22}$ & $9.15\pm^{.55}_{.65}$ & $1.26\pm^{.02}_{.03}$ & $.23\pm_{.02}^{.01}$ & $590\pm^{170}_{170}$ &3.5$ \times 10^{37}$  & $6.1\pm^{.2}_{**} $& $.29\pm^{.29}_{.27} $& $.0054\pm^{.0021}_{.0021}$ & 60 & 1.17\\

706 & $169\pm^{15}_{35}$ & $9.12\pm^{.52}_{.30}$ & $1.31\pm^{.02}_{.05}$ & $.23\pm^{.01}_{.01} $& $486\pm^{120}_{70}$ &3.7 $\times 10^{37}$  & $6.1\pm^{.1}_{**}$ & $.34\pm_{.34}^{.22} $& $.0063\pm_{.0014}^{.0011}$ & 70 & 1.13 \\

707 & $233\pm^{18}_{25}$& $8.49\pm^{.51}_{.20} $& $1.30\pm^{.02}_{.05}$ & $.227\pm^{.016}_{.010}$ & $0.22\pm^{3.0}_{.22}$ &$3.3 \times 10^{37}$ & $6.1\pm^{.1}_{**}$ & $.97\pm^{.15}_{.22} $& $.015\pm^{.03}_{.03}$  & 200 & 1.03\\

\end{tabular}
\end{center}
\caption{The best fit parameter values for the pure thermal Comptonization model.  The error bars represent 90\% confidence levels, while the double asterisks indicate parameters for which the 90\% confidence level is outside the range allowed {\it a priori.}}

\end{table*}

\begin{table*}
\begin{center}
{\bf Best Fit Parameters for the Hybrid Model}

\vskip 1cm
\small
\begin{tabular}{cccccccccccccc}
VP & $kT_{bb}$ & $\ell_{th}$ & $\ell_{nth}$ &$\Gamma$ & $\tau$ & $R$ & $\xi$ & LE & LW & LN & EqW & $\chi^2/\nu$ \\

705 & 192$\pm^{28}_{21}$ & 0.18$\pm^{.29}_{.18}$  & 8.55$\pm^{.45}_{1.55}$ & 3.0$\pm^{.1}_{.10}$ & 1.342$\pm^{.005}_{.015}$ & .233$\pm^{.003}_{.020}$ & 410$\pm^{190}_{130}$  & 6.1$\pm^{.2}_{**}$ & .28$\pm^{.31}_{.28}$ &. 0054$\pm^{.0030}_{.0014}$ & 60 & 1.12\\

706 & 197$\pm^{6}_{10}$ & 0.04$\pm^{.14}_{.04}$  & 8.73$\pm^{.16}_{.13}$ & 3.14$\pm^{.14}_{.15}$ & 1.367$\pm^{.005}_{.005}$ & .236$\pm^{.009}_{.026}$ & 16$\pm^{200}_{2}$  & 6.1$\pm^{.1}_{**}$& 1.0$\pm^{.1}_{.1}$ & .019$\pm^{.003}_{.003}$ & 220 & 1.07\\

707 & 258$\pm^{28}_{34}$ & 6.38$\pm^{1.63}_{3.38}$  & 2.36$\pm^{0.99}_{0.99}$ & 2.1$\pm^{0.5}_{**}$ & 1.31$\pm^{0.02}_{0.06}$ & .237$\pm^{0.016}_{0.027}$ & 5$\pm^{470}_{5}$  & 6.1$\pm^{0.2}_{**}$ & .91$\pm^{0.23}_{0.35}$ & .013$\pm^{.004}_{.003}$ & 180 & 1.04\\

\end{tabular}
\end{center}
\caption{The best fit parameter values for the pure thermal Comptonization model.  The error bars represent 90\% confidence levels, while the double asterisks indicate parameters for which the 90\% confidence level is outside the range allowed {\it a priori.}}

\end{table*}

\label{lastpage} 

\begin{thebibliography}{}

\bibitem{b1} Arnaud, K.A., 1996, in ASP Conf. Series 101, Astronomical Data Analysis Software and Systems V (eds: G.H. Jacoby \& J. Barnes), ASP: San Francisco

\bibitem{b13} Belloni, T. \& Hasinger, G., 1990, A\&A, 227, L33

\bibitem{b20} Beloborodov, A.M., 1999, ApJL, 510, 123

\bibitem{b2} Coppi, P.S., 1998, ``The Physics of Hybrid
Thermal/Non-Thermal Plasmas'' in {\it High Energy Processes in
Accreting Black Holes}, eds. J. Poutanen and R. Svensson, ASP
Conf. Series, Vol. 161, p. 375 (astro-ph/9903158)

\bibitem{b3} Dolan, J.F., 1992, ApJ, 384, 249

\bibitem{b11} Dove, J.B., Wilms, J., Nowak, M.A., Vaughan, B.A., \& Begelman, M.C., 1998, MNRAS, 298, 729

\bibitem{b4} Ebisawa, K., Ueda, Y., Inoue, H., Tanaka, Y, \& White, N.E., 1996,
ApJ, 467, 419

\bibitem{b5} Frontera, F., Palazzi, E., Zdziarski, A.A., Haardt, F., Perola, G.C., Chiapetti, L., Cusumano, G., Dal Fiume, D., Del Sordo, S., Orlandini, M., Parmar, A.N., Piro, L., Santangelo, A., Segreto, A., Treves, A., \& Trifoglio, M., 2001, ApJ, 546, 1027

\bibitem{b14} Ghisellini, G., Haardt, F., \& Fabian, A.C., 1993, MNRAS, 263, L9

\bibitem{b6} Gierlinski, M., Zdziarski, A.A., Done, C., Johnson, W.N., Ebisawa, K., Ueda, Y., Haardt, F., \& Phlips, B., 1997, MNRAS, 288, 958

\bibitem{b7} Gierlinski, M., Zdziarski, A.A., Poutanen, J., Coppi, P.S., Ebisawa, K., \& Johnson, W.N., 1999, MNRAS, 309, 496

\bibitem{b8} Gies, D.R., \& Bolton, C.T., 1986, ApJ, 304, 371

\bibitem{b20} Gilfanov, M., Churazov, E., \& Revnivtsev, M., 1999,
A\&A, 352, 182

\bibitem{b21} Kazanas, D., Hua, X.-M., \& Titarchuk, L., 1997, ApJ, 480,735

\bibitem{b34} Maccarone, T.J., 2001, Ph.D. Thesis, Yale University

\bibitem{b22} Maccarone, T.J., Coppi, P.S., \& Poutanen, J., 2000, ApJL, 537, 107L

\bibitem{b33} Maccarone, T.J., Coppi, P.S. \& Taam, R.E., 1999, BAAS,
195, 3703

\bibitem{b16} McConnell, M.L., Ryan, J.M., Collmar, W., Sch\"onfelder,
V., Steinle, H., Strong, A.W., Bloemen, H., Hermsen, W., Kuiper, L.,
Bennett, K., Phlips, B.F., \& Ling, J.C., 2000, ApJ, 543, 928

\bibitem{b37} Nowak, M.A., Wilms, J. \& Dove, J.B., 2002, MNRAS, in
press (astro-ph/0201383)

\bibitem{b34} Petrucci, P.O., Merloni, A., Fabian, A.C., Haardt, F. \&
Gallo, E., 2001, astro-ph/0108432

\bibitem{b12} Pottschmidt, K., Wilms, J., Nowak, M.A., Pooley, G.G.,
Gleissner, T., Heindl, W., Smith, D.M., \& Staubert, R., 2001,
astro-ph/0202258

\bibitem{b18} Poutanen, J., Krolik, J.H., \& Ryde, F., 1997, MNRAS, 292, L21

\bibitem{b10} Sowers, J.W., Gies, D.R., Bagnulo, W.G.,  Shafter, W., Wiemker, R., \& Wiggs, M.S., 1988, ApJ, 506, 424

\bibitem{b17} Sunyaev, R., \& Tr\"umper, J., 1979, Nature, 279, 506

\bibitem{b19} Young, A.J., Fabian, A.C., Ross, R.R., \& Tanaka, Y.,
2001, MNRAS, 325, 1045

\bibitem{b15} Zdziarski, A.A., Coppi, P.S., \& Lamb, D.Q., 1990, ApJ, 357, 149 

\bibitem{b35} Zdziarski, A.A., Grove, J.E., Poutanen, J., Rao, A.R. \&
Vadawale, S.V., 2001, ApJL, 554, 45

\end{thebibliography}
\end{document}